# Discovery of charge density wave in a correlated kagome lattice antiferromagnet

Xiaokun Teng[1], Lebing Chen[1], Feng Ye[2], Elliott Rosenberg[3], Zhaoyu Liu[3], Jia-Xin Yin[4], Yu-Xiao Jiang[4], Ji Seop Oh[1,5], M. Zahid Hasan[4], Kelly J. Neubauer[1], Bin Gao[1], Yaofeng Xie[1], Makoto Hashimoto[6], Donghui Lu[6], Chris Jozwiak[7], Aaron Bostwick[7], Eli Rotenberg[7], Robert J. Birgeneau[5,8], Jiun-Haw Chu[3], Ming Yi[1], and Pengcheng Dai[1]

[1]Department of Physics and Astronomy, Rice University, Houston, Texas, USA

[2] Neutron Scattering Division, Oak Ridge National Laboratory, Oak Ridge, TN, USA

[3]Department of Physics, University of Washington, Seattle, WA, USA

[4]Laboratory for Topological Quantum Matter and Advanced Spectroscopy, Department of Physics, Princeton University, Princeton, NJ, USA

[5]Department of Physics, University of California, Berkeley, CA, USA

[6]Stanford Synchrotron Radiation Light Source, SLAC National Accelerator Laboratory, Menlo Park, CA, USA

[7]Advanced Light Source, Lawrence Berkeley National Laboratory, Berkeley, CA, USA

[8]Materials Science Division, Lawrence Berkeley National Laboratory, Berkeley, CA, USA

**A hallmark of strongly correlated quantum materials is the rich phase diagram resulting from competing and intertwined phases with nearly degenerate ground state energies[1-3]. A well-known example is the copper oxides, where a charge density wave (CDW) is ordered well above and strongly coupled to the magnetic order to form spin-charge separated stripes that compete with superconductivity[1-3]. Recently, such rich phase diagrams have**

also been revealed in correlated topological materials. In two-dimensional kagome lattice metals consisting of corner-sharing triangles, the geometry of the lattice can produce flat bands with localized electrons[4-11], non-trivial topology[12-14], chiral magnetic order[15,16], superconductivity and CDW order[17-22]. While CDW has been found in weakly electron correlated nonmagnetic $A$V$_3$Sb$_5$ ($A$ = K, Rb, Cs)[17-22], it has not yet been observed in correlated magnetic ordered kagome lattice metals[6,23-28]. Here we report the discovery of CDW within the antiferromagnetic (AFM) ordered phase of kagome lattice FeGe[23-26]. The CDW in FeGe occurs at wavevectors identical to that of $A$V$_3$Sb$_5$[17-22], enhances the AFM ordered moment, and induces an emergent anomalous Hall effect[29,30]. Our findings suggest that CDW in FeGe arises from the combination of electron correlations-driven AFM order and van Hove singularities-driven instability possibly associated with a chiral flux phase[31-38], in stark contrast to strongly correlated copper oxides[1-3] and nickelates[39-41], where the CDW precedes or accompanies the magnetic order.

Charge density wave is a state where electronic charge density in a metal becomes spatially modulated and imposes a new lattice periodicity on a crystal leading to the reorganization of the electronic bands of the parent phase[42]. For materials with weakly correlated electrons such as $A$V$_3$Sb$_5$[22] and Cu$_x$TiSe$_2$[43], CDW can coexist and compete with superconductivity without the involvement of magnetism. In particular, CDW in $A$V$_3$Sb$_5$ is associated with a large anomalous Hall effect (AHE, Fig. 1)[29,30], proposed to arise from Fermi surface nesting of van Hove singularities (vHSs) between the M-L (Fig. 2a) and M-M (Fig. 2b) points of the Brillouin zone (BZ)[44,45], and may host a time reversal symmetry breaking chiral flux phase of circulating currents (Fig. 1d)[31-38]. In strongly correlated electron materials such as copper oxides[1-3] and nickelates[39-41], CDW and associated lattice distortion always appear as a precursor to magnetism, onset at temperatures well above or simultaneously with the magnetic order. There is not yet an example of a coupled CDW and magnetic order where the CDW appears at a temperature well below the magnetic ordering temperature.

In metals with kagome lattices (Fig. 1a), destructive phase interference of nearest neighbor electronic hopping can induce flat electronic bands where the energy scale of electronic correlations exceeds that of the electron kinetic energy, leading to exotic electronic phases[12-14]. When flat bands are near the Fermi level, $E_F$, strong electron correlations can induce magnetic order[6,7,9-11], as seen in kagome lattice FeSn. While magnetism in FeSn orders below $T_N \approx 365$ K with in-plane ferromagnetic (FM) moments in each layer stacked antiferromagnetically along the $c$-axis[27,28], kagome lattice FeGe exhibits collinear $A$-type AFM order with FM moments in each layer parallel to the $c$-axis below $T_N \approx 410$ K (Fig. 1b) and becomes a $c$-axis double cone (canted) AFM structure below $T_{Canting} \approx 60$ K[25,26].

Here we report the discovery of a short-range CDW inside the AFM phase of FeGe below $T_{CDW} \approx 100$ K at wavevectors ($Q$s) identical to that of $A$V$_3$Sb$_5$[17,18,20-22], from a combined study of magnetization (Figs. 1e, 1f, 1i), transport (Figs. 1g, 1h, 1j), neutron scattering (Figs. 2, 3a-h), scanning tunneling microscopy (STM, Figs. 3i,3j)[14], and angle resolved photoemission spectroscopy (ARPES, Fig. 4) experiments. We find that the CDW in FeGe enhances the moment of collinear AFM order (Fig. 1c) and induces an AHE with a magnitude similar to that found in $A$V$_3$Sb$_5$ (Fig. 1k)[29,30], potentially consistent with a chiral flux phase of circulating currents (Fig. 1d)[31-38]. We also observe sets of vHSs near $E_F$ in the electronic band structure separated by the $Q$s identified from neutron diffraction that could be consistent with a nesting origin of the CDW similar to that proposed for $A$V$_3$Sb$_5$[17,18,20-22,44,45]. Since AFM order in FeGe likely arises from electronic correlations of the flat electronic bands near the Fermi level[6,7,9-11], our data suggest that CDW in FeGe is induced by the nesting of vHSs in the spin polarized electronic bands below $T_N$. In contrast to $A$V$_3$Sb$_5$, the CDW order in FeGe is coupled to the AFM order, and provides the first example of a CDW in a magnetically ordered kagome metal.

**Experimental Data**

We first present the transport, magnetization, and overall conclusions of this study. Figure 1e shows temperature versus $c$-axis aligned magnetic field-dependent phase diagram of FeGe as determined from our magnetic susceptibility (Figs. 1f, 1i) and neutron scattering (Figs. 2, 3a-3f) measurements. Below $T_N \approx 410$ K and $T_{Canting} \approx 60$ K, we confirm the collinear $A$-type AFM order (Fig. 1b) and canted AFM structure (Fig. 2c), respectively[25,26]. In addition to the magnetic transitions, temperature dependent susceptibilities (Fig. 1f) for in-plane field along and perpendicular to the $a$- and $b$-axis directions ($\chi_{||a}$ and $\chi_{\perp ac}$) are identical and show a clear reduction around 100 K distinct from both $T_N$ and $T_{Canting}$. While temperature dependent susceptibility along the $c$-axis ($\chi_{||c}$) is consistent with the earlier work[24], the 100 K feature in $\chi_{||a}$ and $\chi_{\perp ac}$ has not been reported. Since we find no change in the magnetic structure across this temperature scale, the reduction of susceptibility is consistent with gapping density of states at the Fermi level by the CDW. The in-plane resistivity ($\rho_{xx}$) exhibits a seemingly featureless metallic behavior (Fig. 1f), yet its temperature derivative shows a clear kink at the same temperature $T_{CDW} \approx 100$ K (Fig. 1g). The $\rho_{xx}$ absolute values change from 150 $\mu\Omega \cdot$ cm at 300 K to ~18 $\mu\Omega \cdot$ cm around 2 K, similar to FeSn and CoSn[10,27] but much larger than the ~30 $\mu\Omega \cdot$ cm at 300 K for $A$V$_3$Sb$_5$[17]. Since the band renormalization factor in FeGe (Fig. 4) is larger than that for $A$V$_3$Sb$_5$[17], the much larger resistivity of FeGe indicates that it's closer to the bad metal regime with stronger electron correlations[46]. The Hall resistivity ($\rho_{xy}$) also shows signatures at $T_{CDW}$. When a magnetic field ($B = \mu_0 H$) is applied along the $c$-axis, a spin-flop transition occurs above a critical field where the $c$-axis aligned moments are flopped to be parallel to the $ab$-plane (Figs. 1e and 1i)[24,47]. Below the spin-flop transition, $\rho_{xy}$ is linear in field (inset in Fig. 1k), which allows us to extract the weak field Hall coefficient simply from $\rho_{xy}/B$ (Fig. 1j). The temperature dependence $\rho_{xy}/B$ drops at 100 K reminiscent of $\chi_{\perp ac}$ and $\chi_{||a}$ (Fig. 1i) and again consistent with the CDW gap formation. At the spin-flop critical field there is a sudden change of both the value and the slope of $\rho_{xy}$ (inset in Fig. 1k), which is a result of both change of the ordinary Hall coefficient due to the spin-flop and the contribution from the AHE, occurring at temperature across $T_{CDW}$. We note that the initial definition of the AHE is the spontaneous Hall effect at zero field[48]. Nevertheless, recent studies have shown that the AHE can also arise in antiferromagnets due to the field induced magnetization and/or non-colinear spin-structure[49]. We extracted the anomalous Hall resistivity ($\rho_{AHE}$) by using the zero-field intercept of the liner fit of $\rho_{xy}$ above the spin flop field as typically done for an antiferromagnet (the inset of Fig. 1k). The $\rho_{AHE}$ is small and fluctuating around zero above $T_{CDW}$. Below the transition, $\rho_{AHE}$ gradually increases as temperature decreases and saturates at a maximum value of 0.027 $\mu\Omega \cdot$ cm. This temperature dependence and the size of $\rho_{AHE}$ is similar to those observed in the $A$V$_3$Sb$_5$ (0.04 $\mu\Omega \cdot$ cm at 5 K)[29,30]. However, unlike $A$V$_3$Sb$_5$ where the AHE is seen as a highly non-linear Hall resistivity at low field ($< 2$ T), there is no sign of AHE in FeGe in the low field range. The fact that the effect of CDW on AHE can only be seen in the spin-flop state where Fe magnetization kicks in (Figs. 1k and 1l) strongly suggests a coupling between magnetism and CDW.

Next, we present our neutron scattering results in the [$H,0,L$] scattering plane. The three-dimensional (3D) and in-plane 2D reciprocal space of FeGe are shown in Figures 2a and 2b, respectively. We start in the normal state where the raw data taken at 440 K above $T_N \approx 410$ K in Fig. 2d show only nuclear Bragg peaks as illustrated in Fig. 2i. On cooling to 140 K, a temperature below $T_N$ but above $T_{CDW}$, a set of new magnetic Bragg peaks at ($H,0,L+0.5$) $Q$s appear (Figs. 2e, 2j). Refinements of the neutron diffraction data confirm the collinear $A$-type AFM structure (Fig. 1b)[25,26]. Upon further cooling to 70 K, a temperature above $T_{Canting}$[25,26], additional peaks appear at ($H+0.5,0,L$), ($H+0.5,0,L+0.5$) (Figs. 2f, 2k), and (0, $K+0.5,L$), (0,$K+0.5,L+0.5$) $Q$s (Figs. 2f, 2k). Since these peaks persist to large $Q$-positions [i.e., at (6.5,0,-5.5)] where the $Fe^{2+}$ magnetic form factor drops off dramatically, they must be nonmagnetic and associated with lattice distortions induced by a CDW (Figs. 2f, 2k). Upon further cooling to below $T_{Canting}$, additional magnetic peaks develop. Figure 2g shows the expanded view of the AFM Bragg peak as marked by the white square in Fig. 2f. We see that a single peak at 70 K clearly splits into three peaks along the $L$ direction (Fig. 2h), consistent with a magnetic structural modulation along the $c$-axis (Figs. 2l, 2m). Our refinements of the single crystal neutron diffraction data at 6 K confirm the canted AFM structure (Fig. 2c)[25,26].

Next, we examine in detail the development of the order parameters across the phase transitions. Figures 3a and 3b show temperature dependence of the (2.5,0,2) and (3.5,0,1.5) CDW peaks across 100 K, respectively. They correspond to $L$=0 and 0.5 along the $c$-axis, similar to that found in $A$V$_3$Sb$_5$[17,18,20-22], and have weak to no hysteresis on warming and cooling processes. The scattering intensity at both $Q$s increase simultaneously below ~100 K, but cannot be fitted with a standard second order phase transition. Figures 3c and 3d show $Q$-scans along the [2.5,0,$L$] and [$H$,0,2] directions, respectively. The scattering is featureless at 140 K but becomes a well-defined peak centered at (2.5,0,2) at 70 K. The peak profile can be best fit by a Lorentzian and is much broader than the instrumental resolution determined by the widths of the nearby nuclear and magnetic Bragg peaks (horizontal bars in Figs. 3c, 3d). Fourier transform of the peak gives correlation lengths of ~33 Å along the in-plane and ~45 Å along the $c$-axis direction, therefore suggesting that this is a short range order. Figure 3e shows temperature dependence of the magnetic ordered moment square obtained by normalizing the integrated intensity of magnetic peak at (2,0,0.5) to moment size squared obtained from refinements of single crystal data at 140 K (Fig. 2e). In addition to confirming $T_N \approx 410$ K (Fig. 3e)[25,26], we find an enhancement of the ordered magnetic moment below $T_{CDW} \approx 100$ K (Fig. 3f), indicating a strong coupling of the magnetic order and CDW. Below ~60 K, incommensurate magnetic Bragg peaks from the canted AFM structure appear, accompanied by a reduction in the magnetic Bragg peak intensity at (2,0,0.5) (Figs. 3f). For comparison, the short-range CDW peak intensity does not change in the canted AFM phase (Figs. 3a and 3b), consistent with the smoothly increasing AHE below ~60 K (Fig. 1k). Therefore, while the CDW order clearly enhances the ordered moment of the collinear $A$-type AFM phase, the CDW intensity does not change on the formation of the canted AFM structure.

To demonstrate that the ~100 K structural superlattice modulation is indeed associated with a CDW transition, we carried out STM measurements on FeGe. Figure 3g reveals atomic lattice of the cryogenic cleaving surface of FeGe. Measuring the differential conductance spectrum that is proportional to the local density of states, we detect a pronounced suppression at the Fermi level, which serves as a candidate of the CDW gap (inset of Fig. 3g) with a size similar to that detected in $A$V$_3$Sb$_5$ systems[18]. While there is weak in-plane superlattice modulations consistent with Fig. 2 from the topographic image, we clearly find the 2 × 2 charge modulation in the differential conductance map at E$_f$ (Fig. 3h and its inset). The wavevectors from our electronic mapping (as marked by the red circles in Fig. 3h inset) match well with the in-plane vector of the CDW detected in Figs. 2 and 3a-3d, supporting the electronic nature of the charge order.

Having confirmed the presence of CDW, we now examine the potential origin of such CDW from the perspective of the electronic structure, as measured by ARPES at 140 K in the AFM state just above $T_{CDW}$. From a detailed photon energy dependence measurement, we have identified 69 eV and 47 eV photons to probe the Γ ($k_z = 0$) and A ($k_z = \pi$) points of the non-magnetic structural Brillouin zone, respectively. The Fermi surfaces probed across both $k_z$ extrema are reminiscent of other kagome metals[6,7,9,44,45], consisting of rounded triangular pockets enclosing small pockets at the K points of the BZ (Fig. 4a). In particular, we identify quasi-straight sections of the Fermi surface connected at the M and L points, similar to those found in $A$V$_3$Sb$_5$[44,45]. More interestingly, we identify two vHSs within proximity of the Fermi level at the M point and one at the L point of the BZ. This can be seen by tracing the dispersions across the M and L points, respectively. Along the direction parallel to the M-K direction, two dispersions are hole-like, where their band top extrapolated from the observable portions of the dispersions traces out electron bands in the orthogonal direction along Γ-M, indicating their saddle-point nature (vHS1 and vHS2 in Fig. 4c). We note that vHS1 is located ~5 meV above Fermi level while vHS2 is ~50 meV below E$_F$. At the L point, we also identify one vHS3 that is similar to vHS1 of the M point, located ~10 meV above E$_F$ (Fig. 4d). Since these features are identified in the state immediately above the onset of the CDW, one can consider a nesting scenario where vHS2 at M below $E_F$ nests with vHS3 at L above $E_F$, giving rise to a $\boldsymbol{Q}_{CDW1}$ of (0.5, 0, 0.5)/(0, 0.5, 0.5), while vHS2 at M below $E_F$ nests with vHS1 at neighboring Ms above $E_F$ to give rise to a $\boldsymbol{Q}_{CDW2}$ of (0.5, 0, 0)/(0, 0.5, 0). The observation of these vHSs and quasi-straight sections of the Fermi surface separated by the $\boldsymbol{Q}$s seen by neutron diffraction would be compatible with a vHS-nesting-driven scenario that has also been proposed for the charge order in the $A$V$_3$Sb$_5$ materials[22].

**Discussion**

It is well established that kagome lattice hosts destructive interference-induced flat bands and vHSs[6,7,9-12]. For FeGe, the location of the kagome-derived flat bands is at $E_F$ in the paramagnetic state[11], where electron correlation effects could drive the system into antiferromagnetically coupled FM planes below $T_N$. Within each FM layer, magnetic order breaks the degeneracy of the electronic bands, splitting the spin-majority and spin-minority electronic bands. The energy splitting of the majority and minority bands increases with decreasing temperature and shifts the electronic bands such that the nesting of vHSs in Fig. 4a to form CDW becomes possible.

Since we observe AHE below the onset of the CDW similar to that seen in $A$V$_3$Sb$_5$[31-38], we speculate that the CDW in FeGe may also be associated with a chiral flux phase of circulating currents as follows (Fig. 1l). At zero field, the chiral circulating currents in each FM kagome plane appearing below the CDW temperature produce *c*-axis aligned and anti-aligned magnetic fields that enhance the iron ordered moment through coupling of their order parameters (left panels of Fig. 1l). Since the circulating currents have opposite directions in two adjacent iron kagome layers, there would be no net magnetization arising from the circulating currents and therefore no observable AHE in zero field. When a $c$-axis field is applied, one would expect two effects to occur that may modify the net magnetization to induce an AHE: i) a spin-flop transition where the iron moment direction changes from along the $c$-axis to mostly the in-plane direction (Fig. 1i, right panels in Fig. 1l); and ii) modulation of the circulating currents in favor of the field direction that would produce a net magnetization. When the Zeeman energy of an applied $c$-axis field is smaller than the spin anisotropy energy that aligns the $c$-axis moment, there would be no spin-flop transition and the system is essentially the same as the zero field state with negligible net magnetization, consistent with no AHE in FeGe in the low field regime.

When Zeeman energy of a $c$-axis field overcomes the spin anisotropy energy at $Q_{AFM}$, a spin-flop transition occurs where the majority of the moments are aligned parallel to the FeGe plane. In principle, the net magnetization induced by the canted iron moments along the $c$-axis can contribute to AHE, but the near zero AHE above $T_{CDW}$ (Figs. 1k and the right panels of Fig. 1l) suggests that the contribution from canted moments angle along the $c$-axis (~2 degree for a 10 T field) is negligible. Within the CDW state, since the in-plane moments no longer lock the circulating currents, the applied $c$-axis field will enhance the circulating currents in the kagome layers with aligned fields and suppress the circulating currents in the layers with anti-aligned fields, therefore producing a net FM moment to induce the observed AHE. As the field-induced spin-flop phase should be maintained below $T_{Canting}$ (right panels of Fig. 1l), we do not expect the zero-field canted AFM phase to affect AHE, in agreement with Fig. 1k. Although we have used chiral flux phase of circulating currents to explain our data, it is clear that electron correlations independent from Fermi surface nesting of vHSs may also explain CDW order and its coupling with magnetism[50]. Regardless of the theoretical explanation of these experimental results, we have identified a kagome metal system in FeGe where strong electron correlations lead to a rich interplay of magnetic order, topology, CDW, and AHE. In the broader context of strongly correlated quantum materials, the CDW that we have found uniquely emerges within and strongly couples to a well-formed magnetic order. This behavior is distinct from that of the stripe phase in cuprates[1-3] and nickelates[39-41] and offers a new platform for exploring emergent phenomena in strongly correlated topological materials.

**Methods:**

**Sample synthesis, structural and composition characterizations.** Polycrystalline FeGe samples were grown by a solid state method[51]. Stoichiometric Fe (99.95% Alfa Aesar) and Ge (99.999%, Alfa Aesar) powders were mixed, ground and pressed into a pellet inside an Argon glovebox. The pellet was loaded inside an $Al_2O_3$ crucible and sealed in an evacuated quartz tube. It was sintered in a box furnace at 1000 °C for 7 days, then at 700 °C for 7 days. Single crystalline FeGe was grown by the chemical vapor transport method[51]. The pellet was ground to powder, then loaded to a sealed quartz tube together with iodine lumps as the transport agent. The temperatures at feedstock and crystallization regions were set to 570 °C and 545 °C, respectively, for 3 weeks. Millimeter size FeGe single crystals were found at the cold end (inset of Extended Data Fig. 1e). Extended Data Fig. 1a shows X-ray Laue pattern of the crystal and Extended Data Fig. 1b plots refinement results comparing calculated and observed Bragg peak intensity. We find no evidence of any other crystal structures, such as $Fe_6Ge_5$ and $Fe_2Ge_3$, as impurity phases[52] because the X-ray and neutron diffraction patterns of these structures[53,54] are quite different from the diffraction patterns shown in Fig. 2d and Extended Data Fig. 1a.

To further confirm that the 100 K feature in Fig. 1f is indeed associated with superlattice distortion, we carried X-ray Laue diffraction experiments at 150 K and 50 K. Extended Data Figs. 1c and 1d show diffraction pattern within the [$H,K$,0.5] zone at 150 K and 50 K, respectively. While there are no observable superlattice peaks at 150 K, superlattice peaks at half integer points appear at 50 K consistent with neutron scattering data of Fig. 2. Since conventional X-ray scattering will not be sensitive to magnetic order, these results further confirm the lattice distortion nature of the superlattice peaks associated with CDW order.

To determine the sample stoichiometry, we carried out elemental analysis using energy-dispersive X-ray (EDX) spectroscopy analysis in a FEI Nano 450 scanning electron microscope on three batches of polished FeGe single crystals. The average stoichiometry of each crystal was determined by examination of multiple points and the outcome suggests that the atomic ratio of Fe:Ge is close to 1:1 (Extended Data Fig. 1e).

To confirm the crystalline quality and stoichiometry of the samples used in our experiments, we perform X-ray single-crystal diffraction experiment at the Rigaku XtaLAB PRO diffractometer housed at the Spallation Neutron Source at ORNL. The measured crystals were suspended in Paratone oil and mounted on a plastic loop attached to a copper pin/goniometer. The single-crystal X-ray diffraction data were collected with molybdenum K$\alpha$ radiation ($\lambda = 0.71073$ Å) at 150 K. More than 4500 diffraction Bragg peaks were collected and refined using Rietveld analysis. The refinement results indicate that Ge2 site and Fe site are full occupied and the occupancy at Ge1 site is 95%. The results suggest that the single crystals are essentially fully stoichiometric with chemical formula: $Fe_{1.00}Ge_{0.98}$.

Table 1. Crystal data and structure refinement for FeGe_S2_150K.

| | |
|---|---|
| Identification code | FeGe_S2_150K |
| Empirical formula | Fe0.33 Ge0.33 |
| Formula weight | 42.81 |
| Temperature | 150(1) K |
| Wavelength | 0.71073 Å |
| Crystal system | Hexagonal |
| Space group | P6/mmm |
| Unit cell dimensions | $a$ = 4.98493(11) Å  $\alpha$ = 90°. |
| | $b$ = 4.98493(11) Å  $\beta$ = 90°. |
| | $c$ = 4.04911(10) Å  $\gamma$ = 120°. |
| Volume | 87.138(4) Å$^3$ |
| Z | 3 |
| Density (calculated) | 2.448 Mg/m$^3$ |
| Absorption coefficient | 12.474 mm$^{-1}$ |
| F(000) | 58 |
| Crystal size | 0.100 x 0.100 x 0.100 mm$^3$ |
| Theta range for data collection | 4.722 to 33.062°. |
| Index ranges | -7<=h<=7, -7<=k<=7, -6<=l<=6 |
| **Reflections collected** | **4577** |
| Independent reflections | 92 [R(int) = 0.0456] |

| | | | | | |
|---|---|---|---|---|---|
| Completeness to theta = 25.242° | | | 100.0 % | | |
| Refinement method | | | Full-matrix least-squares on $F^2$ | | |
| Goodness-of-fit on $F^2$ | | | 1.136 | | |
| Final R indices [I>2sigma(I)] | | | R1 = 0.0346, wR2 = 0.0866 | | |
| R indices (all data) | | | R1 = 0.0346, wR2 = 0.0866 | | |
| Extinction coefficient | | | 0.040(12) | | |

Table 2. Atomic (coordinates x$10^4$) and equivalent isotropic displacement parameters (Å$^2$x $10^3$) for FeGe_S2_150K. U(eq) is defined as one third of the trace of the orthogonalized $U^{ij}$ tensor.

| | x | y | z | Occupancy | U(eq) |
|---|---|---|---|---|---|
| Ge(1) | 0 | 0 | 0 | 0.94 | 6(1) |
| Ge(2) | 3333 | 6667 | 5000 | 1.00 | 4(1) |
| Fe(1) | 5000 | 0 | 0 | 1.00 | 4(1) |

Table 3. Anisotropic displacement parameters (Å$^2$x $10^3$) for FeGe_S2_150K. The anisotropic displacement factor exponent takes the form: $-2\pi^2[ h^2 a^{*2}U^{11} + ... + 2 h k a^* b^* U^{12} ]$

| | $U^{11}$ | $U^{22}$ | $U^{33}$ | $U^{23}$ | $U^{13}$ | $U^{12}$ |
|---|---|---|---|---|---|---|
| Ge(1) | 3(1) | 3(1) | 10(1) | 0 | 0 | 2(1) |
| Ge(2) | 5(1) | 5(1) | 4(1) | 0 | 0 | 2(1) |
| Fe(1) | 3(1) | 3(1) | 4(1) | 0 | 0 | 1(1) |

**Magnetic susceptibility and heat capacity measurements.** Extended Data Figure 2 summarizes the magnetic field dependence of susceptibility measured using a Quantum Design physical property measurement system (PPMS) at Rice. For in-plane applied field, we see increasing susceptibility with increasing field that is weakly temperature dependent across $T_{CDW}$, consistent with canting of the moment along the field direction (Extended Data Figure 2a). When the field is applied along the *c*-axis, spin-flop shown in the inset of Fig. 1e occurs above a critical field, where magnetization increases suddenly due to spin canting. The critical field for spin-flop transition decreases slightly in the CDW phase, as clearly seen in Extended Data Figures 2b and 2c. These results are consistent with earlier work[24] and our inelastic neutron scattering experiments that reveal a decrease in spin anisotropy gap with decreasing temperature below $T_{Canting}$. Extended Data Figures 2d shows temperature dependence of the heat capacity, which confirm that CDW transition is a bulk phase transition.

**Transport Measurements.** Electrical and Magneto-transport measurements were carried out in a 14 T PPMS. Samples were made in a standard four-probe or six-probe contact configuration with current direction in-plane and magnetic field out of the plane (*c*-axis). To eliminate any effects from contact misalignment, the Hall resistivities were symmetrized and anti-symmetrized respectively. Extended Data Figures 3a and 3b show temperature dependent Hall carrier density and Hall mobility calculated from the Hall coefficient and zero field resistivity by assuming a single-band model. Notice that FeGe is a multi-band system. The carrier density and mobility shown here are only effective parameters determined by the carrier density and mobility of each band. Nevertheless, we see a clear reduction in electron density across $T_{CDW}$. Temperature and field dependence of $\rho_{xy}$ and $d\rho_{xy}/dB$ are shown in Extended Data Figures 3c and d, respectively. Extended Data Figure 3e shows temperature dependence of $d\rho_{xy}/dB$ for fields above and below the spin-flop critical field.

Extended Data Figures 4a and 4b show magnetic field dependence of hall resistivity at temperatures across $T_{CDW}$. There is a clear jump in the hall resistivity across the spin-flop transition at temperatures above and below $T_{CDW}$, but the magnitude of the jump becomes smaller with decreasing temperature and changes sign for temperatures approximately below $T_{Canting}$.

The field dependence of Hall resistivity in the $A$-type AFM state is described by $\rho_{xy} = R_1 H$ for $H < H_c$, where $R_1$ is the ordinary Hall coefficient in the AFM state determined by the electronic structure and the scattering rates in the AFM state, $H$ is magnitude of applied field, and $H_c$ is the critical field needed to induce a spin-flop transition. In the field-induced spin-flop state $H > H_c$, we have $\rho_{xy} = R_2 H + \rho_{AHE}$, where $R_2$ is the ordinary Hall coefficient in the spin-flop state determined by the electronic structure and the scattering rates in spin-flop state, $\rho_{AHE}$ is the anomalous Hall resistivity due to the finite magnetization induced by the spin-flop. Therefore, the jump of Hall resistivity at the spin-flop transition in Extended Data Figures 4a and 4b is determined by both the change of ordinary Hall coefficient and the anomalous Hall resistivity.

**Neutron scattering experiments.** Neutron scattering measurements on FeGe were carried out on the CORELLI[55] spectrometer of the Spallation Neutron Source at Oak Ridge National Laboratory, USA. A single crystal of ~30 mg is mounted inside a closed-cycle refrigerator with a base temperature of 6 K and the Mantid package was used for data reduction and analysis. We define the momentum transfer $Q$ in 3D reciprocal space in Å$^{-1}$ as $Q = Ha^* + Kb^* + Lc^*$, where $H$, $K$, and $L$ are Miller indices and $a^* = 2\pi(b \times c)/[a \cdot (b \times c)]$, $b^* = 2\pi(c \times a)/[a \cdot (b \times c)]$, $c^* = 2\pi(a \times b)/[a \cdot (b \times c)]$ with $a = a\hat{x}$, $b = a(\cos 120\, \hat{x} + \sin 120\, \hat{y})$, $c = c\hat{z}$ [Figs. 2a, 2b]. Extended Data Figures 5 shows the raw neutron scattering data within the horizontal [$H$,0,$L$] scattering plane, where broad CDW peaks are seen at $Q_{CDW1} = (0.5,0,0.5)$, $Q_{CDW2} = (0.5,0,0)$. To determine the instrumental resolution and compare with the width of CDW peaks, we show in Extended Data Figures 6 scans along different directions for nuclear Bragg peaks, AFM Bragg peaks, and CDW peaks. We find that CDW peaks are best fit by Lorentzian lineshape, contrasting to typical Gaussian fits for nuclear and magnetic Bragg peaks.

1. **Estimation of canting angle**

In the spin-flop phase, the Zeeman energy generated by a *c*-axis magnetic field is enough to overcome the magnetic anisotropy, yet not enough to overcome exchange interactions between interlayer AFM Fe atoms, hence the net effect will be that the spins turn 90° into the *ab*-plane and have a small canting towards the magnetic field direction. Here, by estimating the order of magnitude of the exchange interaction, we estimate the canting angle after the spin-flop transition under a 10 T field.

Assuming Fe has classical spin $S = 1$, $g = 2$, the Zeeman energy generated by a $B = 10$ T field is $E_Z = g\mu_B SB = 1.15$ meV, which is comparable with the magnetic anisotropy. The AFM $T_N$ of

FeGe is ~ 410 K, corresponding to the inter-planar exchange $J_c = k_B T_N/(gS)$ ~ 17 meV consistent with spin wave measurements. The characteristic canting angle is determined by the ratio between anisotropy and exchange interactions:

$$\theta = \frac{g\mu_B S B}{Z J_c}$$

Using $B$ = 10 T, $J_c$ ~ 17 meV and neighboring atom number $Z$ =2, we get $\theta$ ~ 1.9°.

The canting angle of ~ 2° can be further confirmed by the magnetic moment induced by a 10 T field. The magnetic moment under a 10 T $c$-axis field at 70 K is ~ 4 emu/g = $(4.3 \times 10^{20} \mu_B)$/ $(4.69 \times 10^{21})$ atoms. Therefore, every Fe atom will have ~ 0.09 $\mu_B$ moment, which indicates $gS_z = 0.09$, and the canting angle calculated here is $\arcsin(0.09/2) = 2.6°$, consistent with the estimation from exchange couplings.

2. **Magnetic Form Factor of the Circulating Currents**

The magnetic form factor is the Fourier transform of the magnetic moment in the reciprocal space. In FeGe, there are two types of moment in the CDW flux phase: the moment directly from the magnetic flux induce by circulating charge current (Fig.1d in the main text), or the moment from the Fe ion. The magnetic form factor of the former is determined by the size of the hexagon/triangle in the kagome lattice, while the latter has the form factor of Fe atom itself. Extended Data Figs. 7a-f show the in-plane magnetic form factor generated by the Fourier transform of hexagonal and triangular flux units. Both units have the edge length of $a/2$. Compared with Fe magnetic form factor, the hexagon form factor decays much faster with increasing $|Q|$, while the triangle form factor is comparable to that of Fe. In addition, one would expect that the Fe magnetic form factor to be isotropic in reciprocal space, while the form factor from circulating currents to be highly anisotropic in reciprocal space since the current is confined within the kagome layer.

Extended Data Fig. 7g shows the Q-dependence of the enhanced neutron intensity by subtracting 140 K magnetic peak intensity from 70 K including both in-plane and out-of-plane magnetic Bragg peaks. We compare it with the Q-dependence of the Fe magnetic peaks at 140 K, which should only have Fe moment, and find that they are almost identical. Therefore, the enhanced moment must come from Fe atoms. Additionally, the CDW peak structure factor increases with increasing Q, suggesting that it originates from lattice distortion or modulation (Extended Data Fig. 7h).

## 3. CDW Correlation Lengths

In FeGe, the CDW phase has a smaller correlation length compared with that of magnetic order. As shown in Fig. 2 of the main text, the Bragg peak for the magnetic ordering wavevectors is resolution limited (i.e., has the same FWHM as lattice peaks), suggesting that the magnetic order is long range. On the other hand, the CDW Bragg peak have a larger FWHM when compared with magnetic or lattice peaks (Extended Data Fig. 6). In this section, we calculate the lower limit of the correlation length of CDW in FeGe using inverse Fourier transform of the CDW peaks. Here the word 'lower limit' means we do not consider the peak broadening effect of the instrumental resolution.

The CDW peak is fit by a Lorentzian, $I_q = A \frac{1}{1+(\frac{q-q_0}{\gamma})^2}$, where $A$ is a constant, $q_0$ is the peak center position and $\gamma$ is the HWHM (half width at half maximum, =1/2 FWHM). Its Fourier transform will be $I_x = A\gamma\sqrt{\frac{\pi}{2}}e^{iq_0 x}e^{-\gamma|x|}$. The correlation length ($CL$) here is defined as the FWHM of $|I_x|$, therefore $CL = 2\ln 2/\gamma$. The fitting of the CDW peak gives $\gamma_H = 0.0441$ Å$^{-1}$, $\gamma_L = 0.0273$ Å$^{-1}$ at (2.5,0,2), and gives $\gamma_H = 0.0395$ Å$^{-1}$, $\gamma_L = 0.0296$ Å$^{-1}$ at (3.5,0,1.5). By averaging the width of the two peaks, we get the lower limit of $CL = 33.2$ Å along in-plane directions and $CL = 48.7$ Å along the $c$-axis.

4. **Crystal and Magnetic Structure Refinement Results**

To extract the absolute value of magnetic moment generated by Fe atoms, we performed crystal and magnetic structure refinement using the JANA2006 program[56] for CORELLI data at 300 K, 140 K, and 70 K. Extended Data Figure 8 shows temperature dependence of the scattering map within the [H,K,0.5] plane. The refinement results for magnetic structures are shown in Extended Data Figs. 9a-c. Note that the refinement is only performed on lattice and magnetic peaks at the CDW phase at 70 K, therefore the exclusion of the CDW peaks causes some missing intensity at the lattice Bragg peaks and therefore increases $R_w$ under 70 K. However, the magnetic moment extracted at all three temperatures fits well with the temperature dependence of the order parameter as shown in Fig. 3e in the main text. These results confirm that magnetic structure is the A-type AFM at 300 K, 140 K, and 70 K, and does not change across CDW transition.

As for the incommensurate phase at 6 K, there are two propagation vectors: (0,0,1/2) and (0,0, 0.46). JANA2006 is not able to refine systems with multiple propagation vectors, thus we calculate the magnetic peak intensity using the formula[25,26]:

$$F^2(\boldsymbol{q}) = A\big(f_{\text{Fe}}(\boldsymbol{q})\big)^2 |G_{HKL}|^2 [(1-e_z^2)\delta(\boldsymbol{q}-\tau_{HKL})\cos^2\alpha$$

$$+\frac{1}{4}(1+e_z^2)\delta(\boldsymbol{q}+\boldsymbol{Q}-\tau_{HKL})\delta(\boldsymbol{q}-\boldsymbol{Q}-\tau_{HKL})\sin^2\alpha]$$

where $A$ is a scaling constant, $f_{Fe}(q)$ is the Magnetic form factor of Fe, $G_{HKL}$ is the geometrical structure factor for the *HKL* reflection for the Fe lattice, $|G_{HKL}| = 6$ for even $H$, $|G_{HKL}| = 2$ for odd $H$ in the [*H, 0,L*] plane with half integer or incommensurate *L*. $e_z$ is the projection on the *z* axis of the unit scattering vector $q/|q|$, $\tau_{HKL}$ is the reciprocal lattice vector for the *HKL* reflection ([*H,K,L*] = (integer, integer, half-integer)), $Q = (0,0,0.04)$, and $\alpha$ is the canting angle. Here only the canting angle is a non-trivial variable. Setting $\alpha = 18°$, we calculate the Bragg reflections in the [*H,0,L*] plane as shown in Extended Data Figs. 8d, 8e. The calculated intensity matches well with the experimental data, confirming the correctness of the proposed magnetic structure in the incommensurate phase[25,26].

To understand superlattice peaks induced by the CDW order, we consider a simple model by introducing a small *c*-axis lattice lattice distortion of Fe in an ideal structure of FeGe. The 2 by 2 by 2 superstructure of the CDW phase can be used to simulate the half-integer lattice peak intensity observed at 70 K. Although the full refinement of the superstructure is a challenging task, we can still estimate the extent of lattice distortion induced by CDW using restricted refinements. Assuming that the lattice distortion is on Fe only, and is only along the out-of-plane direction, we can expect superlattice peaks with wavevectors at (1/2,1/2,1/2) and equivalent positions. Under this assumption, we build a FeGe superlattice with $a^* = b^* = 2a = 9.98$ Å and $c^* = 2c = 8.10$ Å, extracted all CDW and integer Bragg peaks from the 70 K data (excluding the magnetic peaks) and multiply the (*H,K,L*) coordinate by a factor of 2. If the CDW phase has the symmetry shown in Fig.1d of the main text, the space group of the superlattice is then reduced to P622 (#177), and the two Fe coordinates are $(1/4,0,1/4+\delta_{c1})$ for Fe1 and $(1/2\ 1/4\ 1/4+\delta_{c2}))$ for Fe2. Fixing all Ge atom positions, we use using the JANA2006 program[56] to refine the two parameters $\delta_{c1}$ and $\delta_{c2}$. The result, as shown in Extended Data Fig. 10, gives $\delta_{c1} = 0.0067$ and $\delta_{c2} = 0.0022$, which corresponds to 0.054 Å and 0.018 Å, respectively. This estimation implies that the Fe atom will have a distortion within ~1 % of the original lattice parameter in the CDW phase.

| Atom | x | y | z |
| --- | --- | --- | --- |
| Fe1 | 0.2500 | 0 | 0.2567 |
| Fe2 | 0.5000 | 0.2500 | 0.2522 |
| Ge1 | 0 | 0 | 0.2500 |
| Ge2 | 0.5000 | 0 | 0.2500 |
| Ge3 | 0.3333 | 0.1667 | 0.5000 |
| Ge4 | 0.3333 | 0.6667 | 0.5000 |
| Ge5 | 0.1667 | 0.3333 | 0.0000 |
| Ge6 | 0.3333 | 0.6667 | 0.0000 |

**STM Measurements.** Single crystals with a size up to 2 mm × 2 mm × 0.3 mm are cleaved mechanically in situ at 77 K in ultra-high vacuum conditions, and then immediately inserted into the microscope head, already at $^4$He base temperature (4.2 K). Topographic images in this work are taken with the tunneling junction set-up V = 60 mV and I = 0.05 nA. Tunneling conductance spectra are obtained with an Ir/Pt tip using standard lock-in amplifier techniques with a lock-in frequency of 977 Hz and a junction set-up of V = 60 mV, I = 1 nA, and a root mean square oscillation voltage of 0.5 mV. Tunneling conductance maps are obtained with a junction set-up of V = 60 mV, I = 0.3 nA, and a root mean square oscillation voltage of 5 mV. For correlated materials, it is not surprising that the CDW order is more often detected in dI/dV maps by STM. Primary examples are cuprate superconductors (4a CDW order is mainly reported by dI/dV maps). We detected weak CDW order taken with bias voltages of -30 meV. We show in Extended Data Fig. 11 a topographic data taken at -30mV and its Fourier transform. Signatures of 2 by 2 charge order are marked by red circles in the Fourier transform data.

**ARPES Measurements.** ARPES measurements were carried out at BL 5-2 of the Stanford Synchrotron Radiation Lightsource and the MAESTRO beamline (7.0.2) of the Advanced Light Source, with a DA30 electron analyzer and a R4000 electron analyzer with deflector mode, respectively. FeGe single crystals were cleaved *in-situ* in ultra-high vacuum with a base pressure better than $5 \times 10^{-11}$ Torr. Energy and angular resolutions used were better than 20 meV and 0.1°, respectively. Extended Data Fig. 12 shows photon energy dependent ARPES measurements.

**Data availability.** The data that support the plots within this paper and other findings of this study are available from the corresponding authors upon reasonable request.

**Acknowledgements**

We thank Di Xiao, Qimiao Si, and Chandan Setty for helpful discussions. The neutron scattering and single crystal synthesis work at Rice was supported by US NSF-DMR-2100741 and by the Robert A. Welch Foundation under Grant No. C-1839, respectively. The ARPES work is supported by the U.S. Department Of Energy (DOE) grant No. DE-SC0021421, the Robert A. Welch Foundation Grant No. C-2024, and the Gordon and Betty Moore Foundation's EPiQS Initiative through grant no. GBMF9470. The transport experiment at the University of Washington is supported by the Air Force Office of Scientific Research under grant FA9550-21-1-0068 and the David and Lucile Packard Foundation. Experimental and theoretical work at Princeton University was supported by the Gordon and Betty Moore Foundation (GBMF4547 and GBMF9461; M.Z.H.), and the US Department of Energy under the Basic Energy Sciences program (grant no. DOE/BES DE-FG-02-05ER46200). The work at the University of California, Berkeley was supported by the U.S. DOE under Contract No. DE-AC02-05-CH11231 within the Quantum Materials Program (KC2202). This research used resources of the Advanced Light Source and the Stanford Synchrotron Radiation Lightsource, both U.S. DOE Office of Science User Facilities under contract Nos. DE-AC02-05CH11231 and AC02-76SF00515, respectively. A portion of this research used resources at the Spallation Neutron Source, a DOE Office of Science User Facility operated by ORNL.

**Author Contributions**

P.D. and M.Y. conceived and managed the project. The single crystal FeGe samples were grown by X.T. and B.G. Neutron scattering and X-ray diffraction experiments were carried out by F.Y. in remote discussion with P.D. and X.T.. Neutron refinements were carried out by X.T., L.C., and K.J.N. Magnetic susceptibility and heat capacity measurements were performed by X.T. and Y.X.. Transport measurements were carried out by E.R., Z.L., J.W.C.. STM measurements are done by J.X.Y., Y.X.J, and M.Z.H.. APRES experiments were done by X.T., J.S.O., R.J.B., and M.Y. with the assistance of M.H., D.L., C.J., A.B., and E.R.. The paper is written by P.D., M.Y., J.W.C., J.X.Y., X.T., and L.C., and all coauthors made comments on the paper.

The authors declare no competing financial interests.

Correspondence and requests for materials should be addressed to M.Y. (mingyi@rice.edu) or P.D. (pdai@rice.edu).

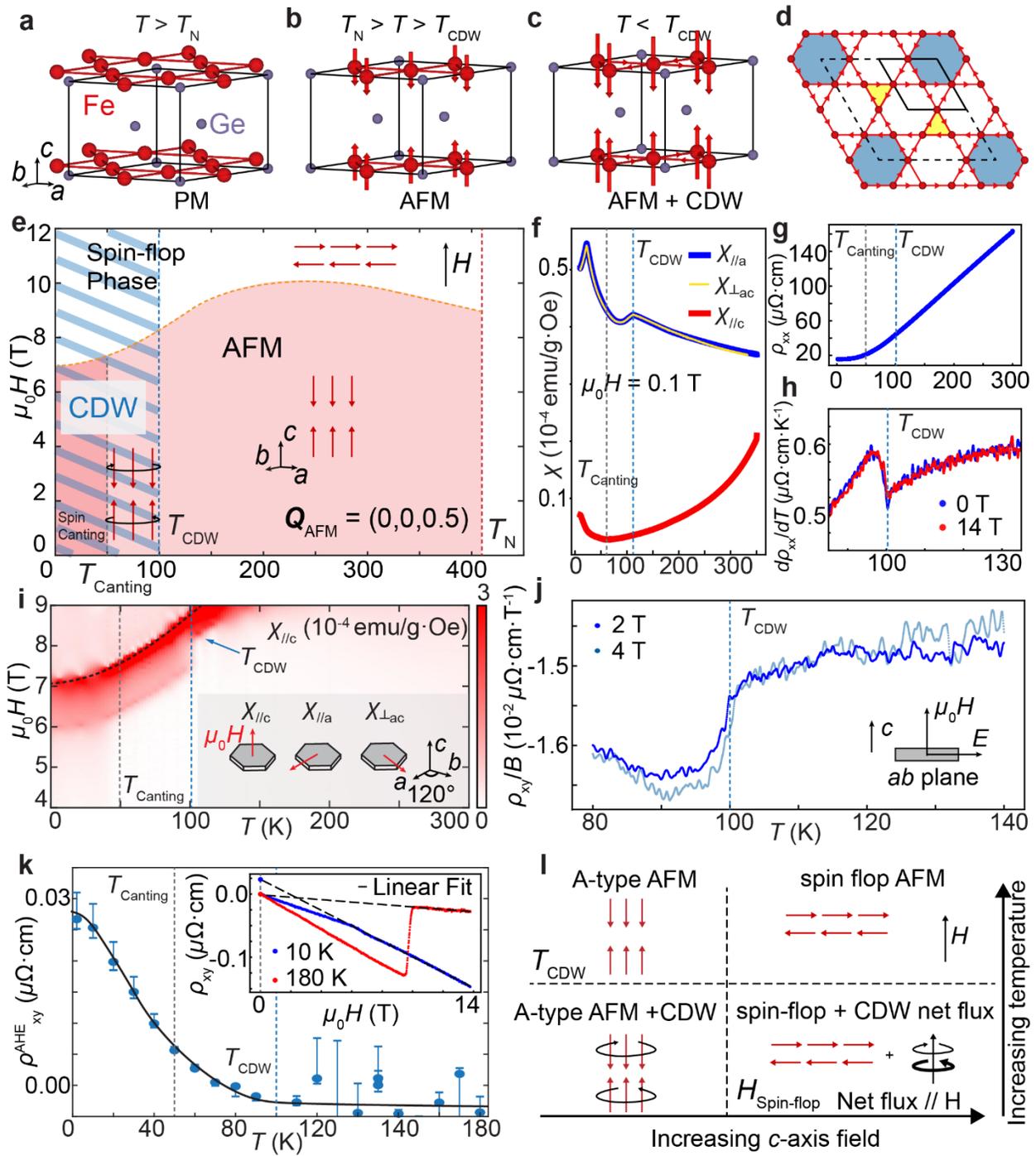

Figure 1. **Crystal and magnetic structure, phase diagram, sample characterization and schematic of flux phase in FeGe. a** Crystal structure in paramagnetic (PM) phase ($T > T_N$). Unit cell is marked with solid lines. The Fe kagome layers are interleaved with Ge layers. **b** Magnetic structure in temperature regime $T_N > T > T_{CDW}$. Spins are aligned within each Fe layer and anti-aligned across the layers. **c** Magnetic structure in $T_{CDW} > T > T_{Canting}$. Moments are enhanced in the CDW state. Red arrows on FeGe bonds indicate possible flux currents. **d** A schematic of the in-plane flux phase. **e** Temperature-field phase diagram of FeGe. Blue shaded area is the CDW state. **f** Magnetic susceptibility of FeGe measured with 0.1 T field. $\chi_{\perp ac}$, $\chi_{||a}$, are measured with field in the *ab* plane, and and $\chi_{||c}$ is measured along the *c*-axis (inset of **i**). There is a kink in $\chi_{\perp ac}$ and $\chi_{||a}$ at $T_{CDW}$. **g,h** Temperature dependence of in-plane resistivity $\rho_{xx}$ and its derivative with respect to temperature. A kink is observed in **h** at $T_{CDW}$. **i** Out of plane magnetic susceptibility at different fields and temperatures. Spin-flop transition temperature is marked with a dashed black line. There is a prominent onset that shows spins are more easily flopped below $T_{CDW}$. **j** Temperature dependence of Hall resistivity with in-plane electric field and magnetic field along the *c*-axis. A decrease in Hall resistivity is observed at $T_{CDW}$. **k** Temperature dependence of $\rho_{xy}^{AHE}$[29,30]. It is extracted by taking the zero field intercept of linear fitting at the spin-flop phase (see inset of **k**). The vertical error bars represent estimation uncertainty in $\rho_{xy}^{AHE}$. **l** Schematic of the interplay between spin-flop and flux phase in FeGe. Below $T_{CDW}$, the in-plane CDW net flux is enhanced selectively after the spin-flop transition.

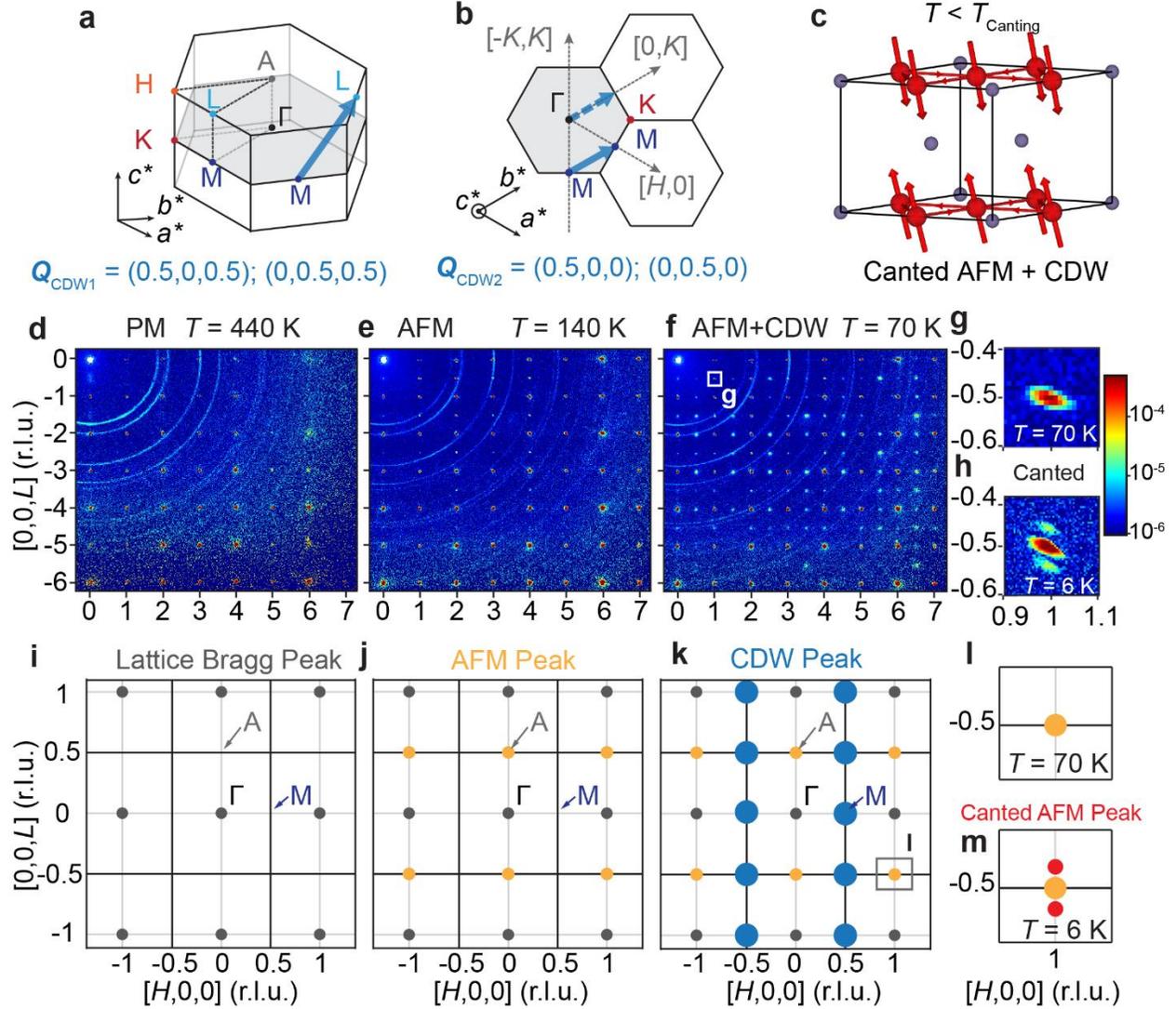

Figure 2 **Brillouin zone, canted AFM structure, neutron diffraction Data and schematic in the [*H*, 0, *L*] plane**. **a,b** 3D and 2D Brillouin zone of FeGe, respectively. The high symmetry points are specified. CDW **Q**s are marked using blue thick arrows. **c** Canted AFM structure of FeGe below $T_{Canting}$. Flux currents in the CDW phase are marked with red arrows on in-plane Fe bonds. **d,e,f,g,h** Neutron diffraction pattern of FeGe at indicated temperatures in the [*H*,0,*L*] plane. The intensity color bar is in log scale. **d** At 440 K, only lattice Bragg peaks at (*H*,0,*L*) ($H, L = 0, \pm 1, \pm 2, \cdots$) are present. **e** At 140 K, the system is in *c*-axis collinear AFM state, AFM

peaks emerge at $\boldsymbol{Q}_{AFM} = (H, 0, L + 0.5)$. **f** At 70 K, CDW coexists with commensurate AFM order, extra CDW peaks appear at $\boldsymbol{Q}_{CDW1} = (H + 0.5, 0, L + 0.5)$ and $\boldsymbol{Q}_{CDW2} = (H + 0.5, 0, L)$. Commensurate Magnetic peak at $\boldsymbol{Q}_{AFM} = (1, 0, 0.5)$ is enlarged in **g**. Below $(H, L$, satellite magnetic peaks emerge, as shown in **h**, indicating additional modulation along the $c$-axis[25,26]. The rings of scatting in **d**, **e**, **f** are from aluminum sample holder. **i,j,k** Schematic of the $[H, 0, L]$ plane reciprocal space maps in the PM phase, AFM phase and AFM+CDW phase. Lattice Bragg peaks, AFM peaks and CDW peaks are shown with gray, orange and blue dots respectively. **l,m** Enlarged magnetic peaks at 70 K and 6 K. Red dots indicate extra incommensurate magnetic peaks below $T_{Canting}$.

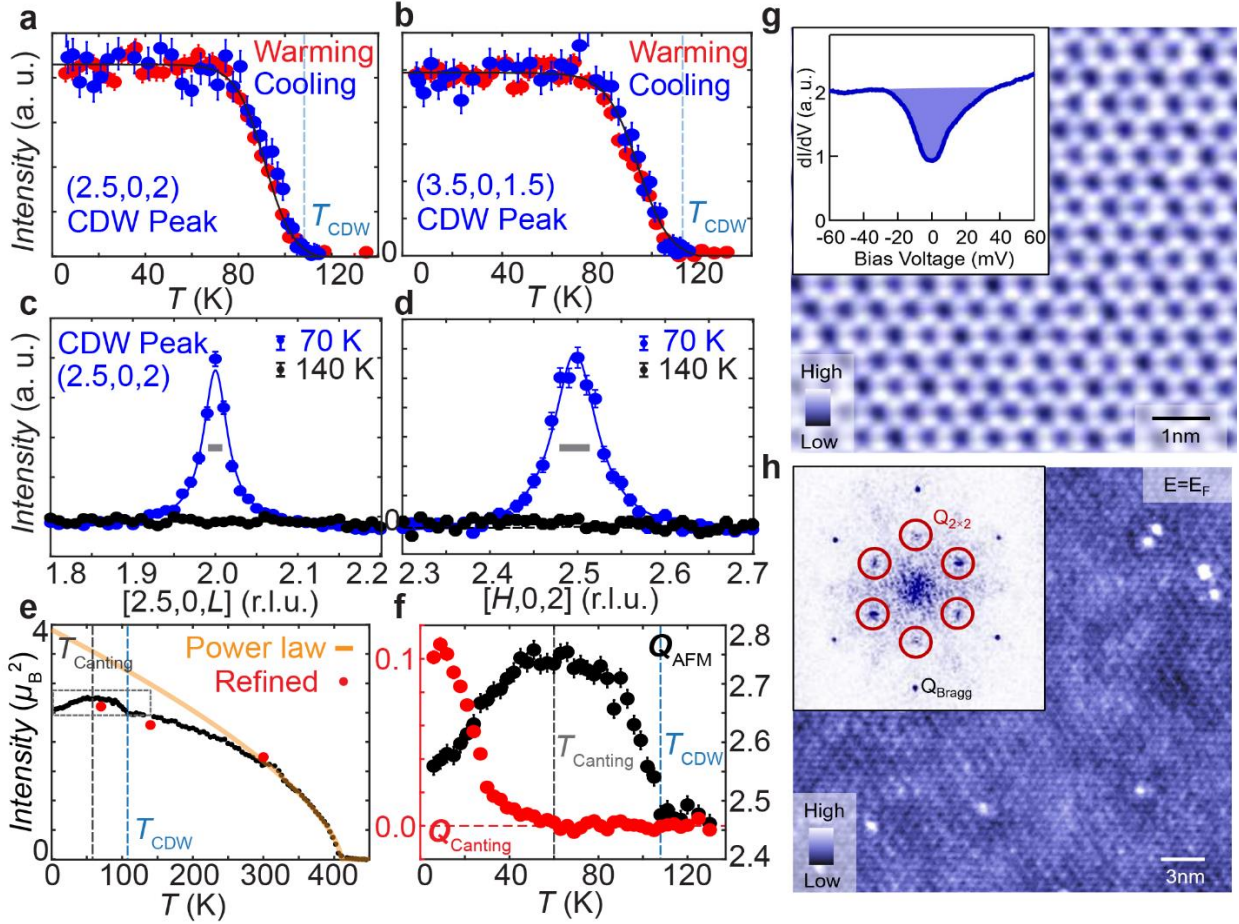

Figure 3 **Order parameter of CDW and magnetic peaks, line-shape of CDW peaks and STM results. a,b** Order parameter of CDW peaks at (2.5, 0, 2) and (3.5, 0, 1.5) respectively. Blue dots show data taken during cooling while red dots show warming. **c,d** Line-shape of the CDW peak at (2.5,0,2) along the $L$ and $H$ directions, respectively. Integration range is $\pm 0.1$ r.l.u. Blue lines are Lorentzian fits of the CDW peaks. Thick gray lines are FWHM of lattice peaks obtained by Gaussian fits. **e** Order parameter of the magnetic peak at $\boldsymbol{Q}_{AFM} = (2,0,0.5)$. Orange line is a power law fit using $I = 3.86(1 - \frac{T}{T_N})^{2\beta}$ with $T_N = 410.3$ K and $\beta = 0.325$. Red dots are refined magnetic moment size. A sudden increase in magnetic moment is observed at $T_{CDW}$. **f,** Low temperature magnetic order parameter. Black and red dots are commensurate and incommensurate magnetic order parameter. At $T_{Canting}$, magnetic Bragg peaks at $\boldsymbol{Q} = (2, 0,$

$0.5\pm\delta$) emerge while intensity of commensurate magnetic peak starts to drop. The vertical error bars in **a-f** are statistical errors of 1 standard deviation. **g,** Atomically resolved topographic image of FeGe with bias voltage $V = 60$ mV and tunneling current $I = 0.5$ nA. The inset shows the differential conductance data, revealing a partially opened energy gap. The tunneling spectrum is taken with $V = 60$ mV, $I = 0.5$ nA, and a bias modulation of 0.5 mV. **h,** Differential conductance map taken at the Fermi level and its Fourier transform (inset). The red circles mark the in-plane $2 \times 2$ charge order vector. The map is taken with $V = 60$ mV, $I = 0.5$ nA, and a bias modulation of 5 mV. All the scanning tunneling microscopy data are taken at 4.2 K.

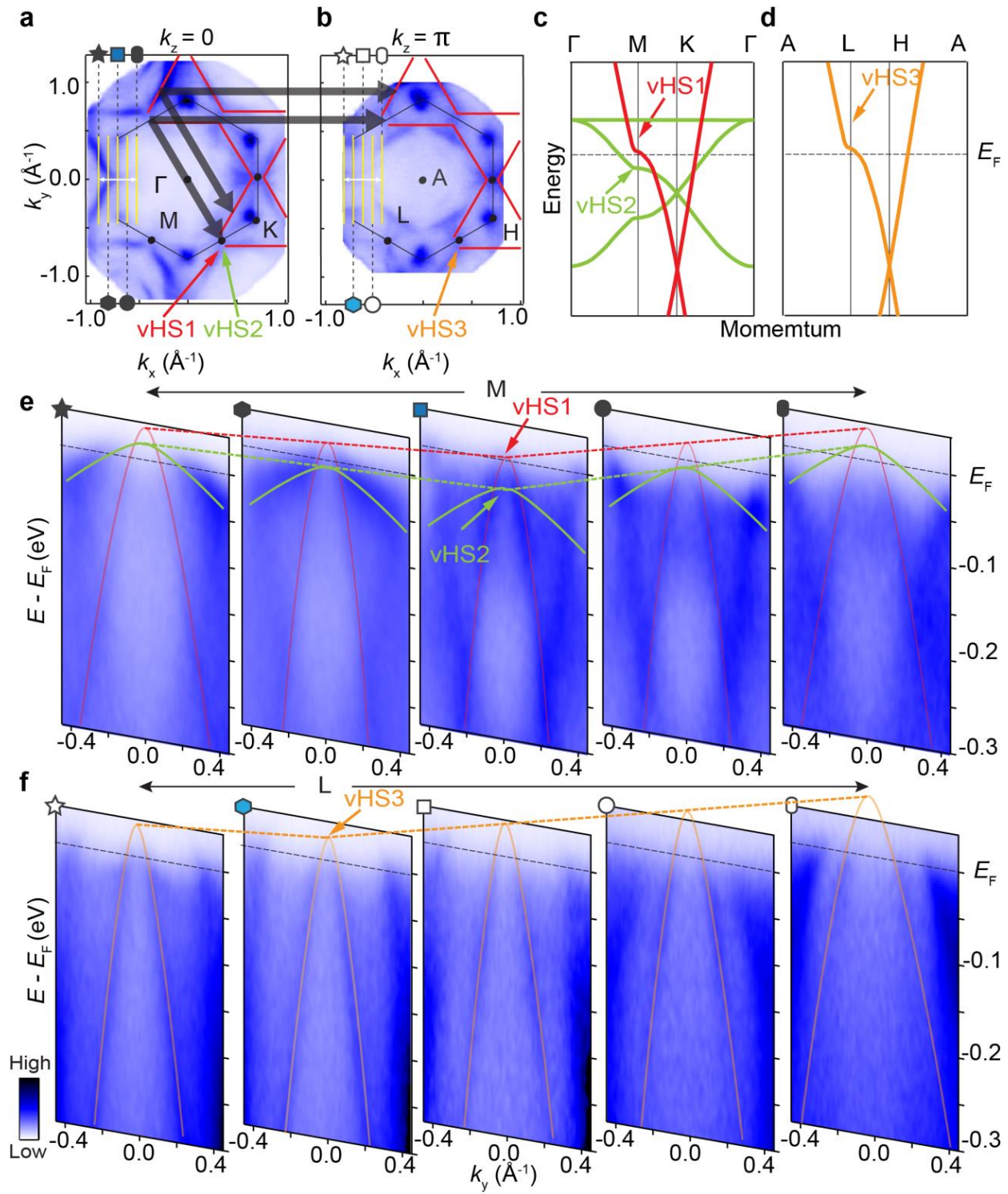

Figure 4 **Schematic and ARPES data of vHSs in AFM FeGe. a,b** APRES measured Fermi surfaces across $k_z = 0$ and $k_z = \pi$ using 69 eV and 47 eV photons, respectively. Polarization used

was linear horizontal. Red lines indicate quasi-straight sections of the Fermi surface connected by $Q_{CDW}$ (gray arrows). Yellow lines mark out each cut in **e** and **f**. **c,d** Schematic of the vHSs in the FeGe electronic structure. **e,f** Dispersions across the vHSs. Solid and dashed lines mark out dispersion of vHSs along $k_y$ and $k_x$ directions, respectively. Cuts in **e** are taken at $k_x$ = -0.93, -0.83, -0.73, -0.63, -0.53 Å$^{-1}$ in the $k_z$ = 0 plane. Blue square indicates the K-M-K cut. Cuts in **f** are taken at $k_x$ = -0.79, -0.73, -0.63, -0.53, -0.43 Å$^{-1}$ in the $k_z$ = $\pi$ plane. Blue hexagon indicates the H-L-H cut. The vHSs are electron-like along Γ-M-Γ (A-L-A) direction and hole-like along K-M-K (H-L-H) direction. There are two vHSs at M point: vHS1 is above the Fermi level while vHS2 is below the Fermi level. One additional vHS3 is located above the Fermi level at the L point. Data are all taken above $T_{CDW}$ at 140 K.